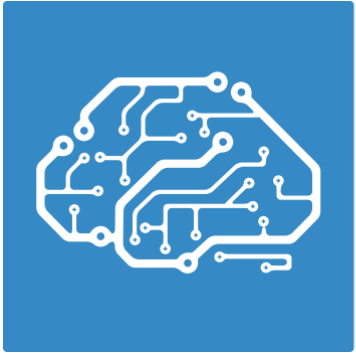

CBMM Memo No. 5    Apr-25-2014

# Sensitivity to Timing and Order in Human Visual Cortex

by

Jedediah M. Singer[1], Joseph R. Madsen[2], William S. Anderson[3], Gabriel Kreiman[1,4,5]

Abstract:
Visual recognition takes a small fraction of a second and relies on the cascade of signals along the ventral visual stream. Given the rapid path through multiple processing steps between photoreceptors and higher visual areas, information must progress from stage to stage very quickly. This rapid progression of information suggests that fine temporal details of the neural response may be important to the brain's encoding of visual signals. We investigated how changes in the relative timing of incoming visual stimulation affect the representation of object information by recording intracranial field potentials along the human ventral visual stream while subjects recognized objects whose parts were presented with varying asynchrony. Visual responses along the ventral stream were sensitive to timing differences between parts as small as 17 ms. In particular, there was a strong dependency on the temporal order of stimulus presentation, even at short asynchronies. This sensitivity to the order of stimulus presentation provides evidence that the brain may use differences in relative timing as a means of representing information.

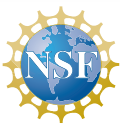

This work was supported by the Center for Brains, Minds and Machines (CBMM), funded by NSF STC award CCF-1231216.

# Sensitivity to Timing and Order in Human Visual Cortex

Jedediah M. Singer[1], Joseph R. Madsen[2], William S. Anderson[3], Gabriel Kreiman[1,4,5]

[1]Department of Ophthalmology, Boston Children's Hospital, Harvard Medical School, Boston, MA 02115, USA
[2]Department of Neurosurgery, Boston Children's Hospital, Harvard Medical School, Boston, MA 02115, USA
[3]Department of Neurosurgery, Johns Hopkins Hospital, Johns Hopkins University School of Medicine, Baltimore, MD 21204
[4]Center for Brain Science, Harvard University, Cambridge, MA 02138, USA
[5]Swartz Center for Theoretical Neuroscience, Harvard University, Cambridge, MA 02138, USA

Correspondence: Jedediah.Singer@childrens.harvard.edu, (617) 919-2246
Jedediah Singer
3 Blackfan Circle, Room 13075
Boston, MA 02215

## Acknowledgments

Financial support was provided by NIH and NSF. The authors would like to thank Roozbeh Kiani, Hanlin Tang, Radhika Madhavan, Kendra Burbank, John Maunsell, and Chou Hung for comments on the manuscript; Morgan King for the stimuli; the patients for their participation in this study; and Sheryl Manganaro and Lixia Gao for facilitating data collection. The authors declare no competing financial interests.

**Abstract**

Visual recognition takes a small fraction of a second and relies on the cascade of signals along the ventral visual stream. Given the rapid path through multiple processing steps between photoreceptors and higher visual areas, information must progress from stage to stage very quickly. This rapid progression of information suggests that fine temporal details of the neural response may be important to the brain's encoding of visual signals. We investigated how changes in the relative timing of incoming visual stimulation affect the representation of object information by recording intracranial field potentials along the human ventral visual stream while subjects recognized objects whose parts were presented with varying asynchrony. Visual responses along the ventral stream were sensitive to timing differences between parts as small as 17 ms. In particular, there was a strong dependency on the temporal order of stimulus presentation, even at short asynchronies. This sensitivity to the order of stimulus presentation provides evidence that the brain may use differences in relative timing as a means of representing information.

## Introduction

Shape recognition is essential for most visual tasks and depends on continuous integration of visual cues over space and time. Shape recognition relies on the semi-hierarchical cascade of linear and non-linear steps along the ventral visual stream (Haxby et al., 1991; Rolls, 1991; Logothetis and Sheinberg, 1996; Tanaka, 1996; Connor et al., 2007). Several studies have documented the spatial integration properties of neurons along the ventral stream, showing that receptive field sizes increase from early visual cortex all the way to inferior temporal cortex (ITC) (Gattass et al., 1981; Gattass et al., 1988; Kobatake and Tanaka, 1994; DiCarlo and Maunsell, 2000; Yoshor et al., 2007; Dumoulin and Wandell, 2008; Agam et al., 2010). Further, the presence of multiple objects can significantly influence the physiological responses within the receptive field (Missal et al., 1999; Gawne and Martin, 2002; Zoccolan et al., 2007; Agam et al., 2010; Baeck et al., 2013).

Comparatively less is known about the dynamics underlying temporal integration of visual information, particularly in the highest echelons of ventral cortex. Mean response latencies progressively increase along the ventral stream by ~15 ms at each stage (Schmolesky et al., 1998), and selective responses to complex shapes have been reported in ITC at 100-150 ms post-stimulus onset both in macaque monkeys (Richmond et al., 1990; Keysers et al., 2001; Hung et al., 2005) and humans (Thorpe et al., 1996; Liu et al., 2009). This rapid progression in information transmission suggests that fine temporal details of the neural response may be important to the brain's encoding of visual signals and has led to theories describing visual recognition via bottom-up and hierarchical concatenation of linear and non-linear processing steps (Fukushima, 1980; Wallis and Rolls, 1997; Riesenhuber and Poggio, 1999; Deco and Rolls, 2004; Serre et al., 2007). At the same time, neurons often show response durations that span several tens to hundreds of milliseconds (Richmond et al., 1990; Ringach et al., 1997; Keysers and Perrett, 2002; De Baene et al., 2007), which may endow them with the potential to integrate visual inputs over time. Behavioral studies have suggested windows of temporal integration that range from several tens to hundreds of ms (Clifford et al., 2004; Singer and Sheinberg, 2006; Anaki et al., 2007; Schyns et al., 2007).

To further our understanding of how sensory stimuli are integrated over space and time in the human ventral visual stream, here we investigated whether changes in the relative timing of incoming visual stimulation affect the representation of object information. We recorded intracranial field potentials while subjects recognized objects whose parts were presented with varying asynchrony and used quantitative models to describe the extent of spatiotemporal integration by the ensuing physiological signals. Temporal asynchrony as short as 17 ms led to significant differences in the physiological responses. Furthermore, distinct responses were evoked by altering the relative order of object part presentation. These results demonstrate that the human ventral visual stream is sensitive to relative timing on scales of approximately 17 ms.

## Materials and Methods
### Subjects

Subjects were 5 patients (4 female) at Boston Children's Hospital and Johns Hopkins Medical Center, with subdural and/or depth electrodes (Ad-Tech, Racine, WI) implanted for clinical purposes as part of epilepsy treatment. The number of electrodes as well as their location was exclusively dictated by clinical considerations. The number of electrodes per patient ranged from 84 to 186 (total = 628); the electrode locations are described in Table 1. Sixteen healthy subjects (10 female) performed a psychophysics experiment described below. All procedures were performed with informed consent and approved by the Boston Children's Hospital and Johns Hopkins Medical Center Institutional Review Boards.

**Task**

Subjects were shown asynchronously presented image parts, and were asked to identify the images (Figure 1A). Each subject saw grayscale images with flattened intensity histograms from one of two sets of four stimuli; each stimulus was constructed of two parts (a top part and a bottom part), and each part was present in two images (Figure 1B). Subjects were familiarized with the images and their names before the experiment. It was made explicit that sometimes the two parts of the image would be shown at different times, but that the subject should still respond according to the image whose parts were presented. A fixation cross persisted throughout the trial. Each image subtended approximately 5-6 degrees of visual angle vertically and 4-6 degrees horizontally. Each trial began with 500 ms of low-contrast phase-scrambled noise at 60 Hz. One of the two image parts then appeared on the screen for one screen refresh (17 ms). The second image part was presented, also for 17 ms, with a stimulus onset asynchrony (SOA) of 0 (both parts appeared simultaneously), 1, 2, 3, 6, 15, or 42 screen refreshes (0, 17, 33, 50, 100, 250, or 700 ms, respectively). The flickering noise continued behind and between the two parts, and for 500 ms after the onset of the second image part. The subject was then presented with a screen giving the four image choices and the corresponding buttons to press (4-alternative forced choice). Mapping between images and button presses remained fixed throughout the experiment. No correct/incorrect feedback was provided, except for an overall score at the end of each block of 40 trials. The order of presentation of the different images and SOA values was randomized.

**Physiological recordings**

Electrophysiological data were recorded and digitized at 500 Hz, 1000 Hz, 1024 Hz, or 2000 Hz (depending on the subject) using either an XLTEK or Nihon Kohden clinical system. All analyses of electrophysiological data were performed with MATLAB software (Mathworks, Natick, MA). We subtracted the mean across all electrodes from each channel to reduce externally induced artifacts. We bandpass filtered the data between 1 Hz and 100 Hz, with a notch filter at 60 Hz to remove line noise. To reduce artifacts, we excluded trials in which any sample was more than 4 standard deviations (over all trials) from the mean response (over trials with the same image and SOA). This excluded 4.5%, 4.3%, 2.0%, 5.9%, and 3.7% of trials, respectively, for the five subjects.

Electrodes were localized by co-registering a pre-operative structural MRI scan with a post-implantation CT scan. We used Freesurfer software to compute a 3D representation of the cortical surface from the structural MRI, and manually located each electrode shown in the CT scan on this surface. Brain regions and Talairach coordinates were also calculated using Freesurfer (Fischl et al., 2004; Destrieux et al., 2010). We excluded 5 electrodes because they appeared to be shorted.

**Data Analyses**

*Visual responsiveness*

To assess visual responsiveness, we computed the mean across all whole-image trials for each electrode. An electrode was considered to be visually responsive to an image if the range

(max-min) of the mean response between 50 ms and 350 ms after image onset was larger than chance, as determined by a permutation test in which responses to individual trials were randomly multiplied by either 1 or -1 (10,000 iterations, p<0.0001).

We also used a permutation test to partition visual responses into those that showed order sensitivity and those that did not. Trials at 17 ms SOA in both orders were randomly partitioned into two groups; the root-mean-square error (RMSE) between the means of the two groups was calculated (between 50 ms and 350 ms after image onset), and this process was repeated 5000 times. The actual RMSE between mean responses was then compared to this distribution. We used the same kind of permutation test to evaluate differences between responses to whole images and different SOA values (in both orders).

In some cases, responses to two-part stimuli were dominated by the response to one of the parts; while this sometimes reflected a winner-take-all interaction between two part responses, in other cases the weaker part did not elicit a response even in isolation. These latter responses likely reflected limited receptive fields (spatial or object-related) and thus were not pertinent to this study of interactions between parts. We therefore excluded all responses in which asynchronous presentations could be described by the response to one of the parts (Eq. 2, below, with probability p>0.05) *and* one of the responses to individual parts failed to reach a threshold (p<0.001) in the same type of permutation test used to ascertain responsiveness to whole images.

## *Quantitative models to describe the interactions of image parts*

We evaluated how well we could explain the responses at different SOAs using models that combine the responses to the two image parts. Let $t$ be the time from onset of the first image part. Let $r_B(t)$ denote the response (averaged across trials) to the bottom image part (calculated from the SOA=700 ms trials), and similarly let $r_T(t)$ be the response to the top part. Let $r_{TB}(t,SOA)$ denote the response at time $t$ (averaged across trials) to presentation of both parts, with the top part appearing SOA ms before the bottom part (*SOA>0* represents trials where the top part was presented first and *SOA<0* represents trials where the bottom part was presented first). To focus on the most critical time window for visual responses and interactions between responses to image parts, all the model evaluations were performed between $t \geq 50$ ms and $t \leq 350$ ms. All parameters were optimized using the method of least squares. We denote by $\hat{r}_n(t,SOA)$ the prediction for model $n$ (defined below) at a given time $t$ and SOA. We restricted the evaluation of models to $17 \leq |SOA| \leq 50$ ms because independent responses to the two parts were apparent at SOA=100 ms and 250 ms.

The first model evaluated whether asynchronous presentation had any effect on the response, by describing the responses to SOA≠0 ms using the responses to the whole images (SOA=0 ms, "whole image model"):

$$\hat{r}_1(t,SOA) = r_{TB}(t,0) \qquad \text{Eq. 1}$$

This model had 0 free parameters.

We then considered whether the response to either individual part alone could describe the responses to the combined image parts ("best part model"). For SOA≤0, we defined $\hat{r}_{2B}(t,SOA) = r_B(t)$, $\hat{r}_{2T}(t,SOA) = r_T(t-SOA)$. When SOA>0, $\hat{r}_{2B}(t,SOA) = r_B(t+SOA)$ and $\hat{r}_{2T}(t,SOA) = r_T(t)$ and then:

$$\hat{r}_2(t,SOA) = \arg\max_p [\hat{r}_{2B}(t,SOA), \hat{r}_{2T}(t,SOA)] \qquad \text{Eq. 2}$$

where $p$ is the probability of $\hat{r}_{2B}(t,SOA)$ or $\hat{r}_{2T}(t,SOA)$ (defined below, calculated across $17 \leq |SOA| \leq 50$). In other words, this model assumes that one of the two parts is exclusively responsible for driving the overall response regardless of SOA. This model had 0 free parameters.

We next considered a model where the two parts could be linearly combined ("simple linear model"):

$$\hat{r}_3(t,SOA) = \begin{cases} c_B r_B(t) + c_T r(t-SOA) & SOA \leq 0 \\ c_B r_B(t+SOA) + c_T r_T(t) & SOA > 0 \end{cases} \qquad \textbf{Eq. 3}$$

This model had two free parameters ($c_B$, $c_T$). Note that here $c_B$ and $c_T$ do not depend on SOA.

While the simple linear model was able to describe the combined responses in many of the electrodes (see text), we noted in many cases that the contribution of each part seemed to depend on the order in which the two parts were presented. We therefore extended the simple linear model to allow a different contribution for each possible SOA value and order ("general linear model"):

$$\hat{r}_4(t,SOA) = \begin{cases} c_B(SOA)r_B(t) + c_T(SOA)r_T(t-SOA) & SOA < 0 \\ c_B(SOA)r_B(t+SOA) + c_T(SOA)r_T(t) & SOA > 0 \end{cases} \qquad \textbf{Eq. 4}$$

This model had 2 free parameters per SOA. Hence, the most general version included 22 parameters. To compare against the equations above, we considered a restricted version of Eq. 4 with 12 parameters, two per SOA. Note that Eqs. 2-3 are special cases of Eq. 4.

We also considered an alternative family of models that, rather than predicting responses as linear combinations of responses to parts, described the responses at longer SOA values based on the responses at the shortest SOAs. These models aimed to capture invariances across SOA values. Since they were based upon responses at ±17 ms SOA, they were defined only at SOAs of ±33 ms and ±50 ms. When comparisons were made between these models and the previous models (Eq. 1-4), those models were re-calculated using only these SOA values.

Equating the number of fit parameters to the simple linear model (Eq. 3), we fit one coefficient to the response in each order ("short SOA model with order"):

$$\hat{r}_5(t,SOA) = \begin{cases} k_1 r_{TB}(t,-17ms) & SOA < -17ms \\ k_2 r_{TB}(t,17ms) & SOA > 17ms \end{cases} \qquad \textbf{Eq. 5}$$

where the parameters were $k_1$ and $k_2$.

In conceptual parallel to the general linear model in Eq. 4, we extended Eq. 5 by allowing a parameter for each SOA value ("general short SOA model"):

$$\hat{r}_6(t,SOA) = \begin{cases} k(SOA)r_{TB}(t,-17ms) & SOA < -17ms \\ k(SOA)r_{TB}(t,17ms) & SOA > 17ms \end{cases} \qquad \textbf{Eq. 6}$$

with parameters $k(-50 ms)$, $k(-33 ms)$, $k(50 ms)$, $k(33 ms)$. Note that, because there was only one parameter for each SOA value, this model had half the number of parameters of the general linear model (Eq. 4).

The presentation of different parts in asynchronous fashion generates spatiotemporal features that are absent when the parts are presented simultaneously. These features may lead to electrophysiological responses and percepts that differ between the two presentation orders irrespective of the part shapes. To examine whether such shape-independent spatiotemporal features could explain the physiological responses, we considered a model where the two *other* parts were presented in the same order and with the same SOA. For example, the response to w6 with t3 followed 17 ms later by b3 was compared with the response to the w7 with t4 followed 17 ms later by b4. This model can be described by:

$$\hat{r}_7(t,SOA) = r_{T'B'}(t,SOA) \qquad \textbf{Eq. 7}$$

where *T'* and *B'* indicate the top and bottom parts *not* used in the response being modeled.

## *Model Evaluation*

We calculated the probability that the data could be described by each equation to evaluate each of the model fits. We used a permutation test based on the distribution of differences between mean responses generated by identical stimuli. To evaluate how well model *n* described the data for a given SOA, we first computed the root-mean-square error between the model and the actual responses:

$$\text{RMSE}_n(SOA) = \sqrt{\frac{\sum_{t=50}^{350}(r_{TB}(t,SOA) - \hat{r}_n(t,SOA))^2}{(350-50)s/1000}} \qquad \textbf{Eq. 8}$$

where *s* is the sampling rate in samples/sec and the sum in the numerator includes all values of *t* from 50 to 350 ms for which there were data samples. $\text{RMSE}_n(SOA)$ takes the value of 0 for a perfect fit and is bounded above by the variation in $r_{TB}$ over time.

We compared the difference between the model fit and the data (evaluated by $RMSE_n$) against the trial-to-trial variability observed at the electrode in question in repeated presentations of the whole image. To make this comparison, we estimated the distribution of RMSEs between the averages of complementary subsets of responses to the whole image. Let $r_{i,TB}(t,0)$ indicate the response at time *t* in presentation *i* (*i*=1,…,*N*) of the whole image; note that

$r_{TB}(t,0) = \frac{1}{N}\sum_{i=1}^{N} r_{i,TB}(t,0)$. We consider two non-overlapping equal sized random partitions of the

*N* trials: $\pi_1$ and $\pi_2$ ($\pi_1 \cup \pi_2 = \{1,...,N\}$, $\pi_1 \cap \pi_2 = \emptyset$ and $|\pi_1| = |\pi_2|$). We define the average

response over each partition $_1r_{TB}(t,0) = \frac{1}{|\pi_1|}\sum_{i \in \pi_1} r_{i,TB}(t,0)$ and $_2r_{TB}(t,0) = \frac{1}{|\pi_2|}\sum_{i \in \pi_2} r_{i,TB}(t,0)$ and the

RMSE between those two mean responses: $\text{RMSE}_{whole} = \sqrt{\frac{\sum_{t=50}^{350}(_1r_{TB}(t,0) - _2r_{TB}(t,0))^2}{(350-50)s/1000}}$

This procedure was repeated 5000 times (indexed by *j*) to generate a sampling distribution *D* from $\{RMSE_{whole,j}\}$. We compared the value of $\text{RMSE}_n(SOA)$ for a given model against the distribution *D*. Let $p_{SOA}$ be the percentile of $\text{RMSE}_n(SOA)$ with respect to *D*:

$$p_{SOA} = \frac{\left|\{RMSE_{whole,j} | RMSE_{whole,j} \leq RMSE_n(SOA)\}\right|}{5000} \qquad \textbf{Eq. 9}$$

where || denotes the cardinality of the set and $0 < p_{SOA} < 1$. A large value of $p_{SOA}$ indicates a poor model, while $p_{SOA} = 0.5$ indicates that the difference between the model and the actual data is as small as the median difference between different partitions of the responses to the whole image—in other words, the model fits the data approximately as well as can be expected given

trial-to-trial variability. We then obtained the probability *p* that the data at *even the worst-case SOA* could be generated by the model in question:

$$p = \min_{SOA}(1 - p_{SOA}).$$  **Eq. 10**

In other words, a model's effectiveness was evaluated based on its worst performance across all SOAs. To prevent bias, these probabilities were only compared for models fit and evaluated on the same set of SOA values. Note also that parameters were fit based on minimization of overall squared error across all considered SOA values, but evaluated based on the error at the worst-fit SOA value.

*Akaike Information Criterion*

To compare models with different numbers of parameters, we calculated the corrected Akaike Information Criterion (AICc) (Akaike, 1974; Burnham and Anderson, 2002). Differences in this value describe how much more likely one model is than another to minimize the information lost due to replacing the data with the models. When calculating AICc, we estimated a model's likelihood as $\prod_{SOA}(1 - p_{SOA})$. We set a threshold of 4 for deeming one model to be significantly better than another, which corresponds to a likelihood ratio of approximately 7.4.

*Order Tuning Index*

Often, responses appeared to change when part order changed, but remained relatively constant in the face of other manipulations of SOA (see text). To quantify this observation, we computed a symmetric comparison matrix (Figure 4) that described the differences in the responses for any two SOA values, $SOA_1$ and $SOA_2$:

$$RMSE(SOA_1, SOA_2) = \sqrt{\frac{\sum_{t=50}^{350}(r_{TB}(t, SOA_1) - r_{TB}(t, SOA_2))^2}{(350-50)s/1000}}.$$  **Eq. 11**

Note that when (and only when) averaging across responses to plot Figure 4D, F we normalized these values for each response by dividing them by the median of the distribution *D* of $RMSE_{whole}$ (see above) for that response. We then defined the across-order average difference,

$$RMSE_{across} = \frac{1}{16} \sum_{\substack{SOA_1=-100 \\ SOA_1 \neq 0}}^{SOA_1=100} \sum_{\substack{SOA_2=-100, SOA_2 \neq 0 \\ sign(SOA_2) \neq sign(SOA_1)}}^{SOA_2=100} RMSE(SOA_1, SOA_2)$$  **Eq. 12**

the within-order average difference,

$$RMSE_{within} = \frac{1}{12} \sum_{\substack{SOA_1=-100 \\ SOA_1 \neq 0}}^{SOA_1=100} \sum_{\substack{SOA_2=-100, SOA_2 \neq SOA_1 \\ sign(SOA_2) = sign(SOA_1)}}^{SOA_2=100} RMSE(SOA_1, SOA_2)$$  **Eq. 13**

and defined the order tuning index (OTI):

$$OTI = \frac{RMSE_{across} - RMSE_{within}}{RMSE_{across} + RMSE_{within}}$$  **Eq. 14**

OTI takes a value > 0 when the differences across presentation orders are larger than the differences within the same order and is negative when the differences within orders are larger (OTI ranges from -1 to 1).

In the definition of OTI above, the absolute differences in the SOA values in the *across* conditions are larger than the ones in the *within* conditions. This does not actually reflect a larger

SOA, but does reflect that the temporal positions of the top and bottom parts have been reversed. To ensure that OTI values would not be overly biased in a positive direction by these differences in SOA values, we considered the following alternative definition, which limited the *across* condition to short SOAs:

$$RMSE'_{across} = \tfrac{1}{4} \sum_{\substack{SOA_1=-33 \\ SOA_1 \neq 0}}^{SOA_1=33} \sum_{\substack{SOA_2=-33, SOA_2 \neq 0 \\ sign(SOA_2) \neq sign(SOA_1)}}^{SOA_2=33} RMSE(SOA_1, SOA_2) \qquad \text{Eq. 15}$$

$$OTI' = \frac{RMSE'_{across} - RMSE_{within}}{RMSE'_{across} + RMSE_{within}} \qquad \text{Eq. 16}$$

We repeated all calculations leading up to OTI and OTI', including determinations of visual responsiveness and order sensitivity, using a temporal window from 50 to 200 ms after stimulus onset.

**Psychophysics task**

We conducted a variation of the main task to evaluate whether it is possible to detect stimulus asynchrony under the same stimulus presentation conditions. In this psychophysics experiment, 16 healthy subjects (10 female) were asked to indicate whether the two parts were presented simultaneously or asynchronously (two-alternative forced-choice task). Eye position was tracked using an Eyelink 1000 Remote system, and trials were initiated by 500 ms fixation within 3 degrees of the fixation point at the center of the screen. Each subject completed 5 blocks of 48 trials. In half of the trials, the two parts were presented simultaneously, and in the rest the two parts were presented with 17, 33, 50, and 100 ms SOA.

**Results**

Subjects identified which of four possible images was shown in a task where the two constituent image parts were briefly flashed (17 ms) with a stimulus onset asynchrony (SOA) ranging from 0 ms (whole images) to 700 ms (Figure 1). Subjects could identify the images in spite of the rapid presentation and intervening frames of visual noise (mean performance = 76% correct, chance = 25%); performance did not depend on SOA for any subject (chi-square tests, $p>0.1$). We recorded physiological responses from 628 subdural and depth electrodes in 5 subjects during the task. The responses at an example electrode located in the left fusiform gyrus are shown in Figure 2. Consistent with previous neuroimaging work (Haxby et al., 2001; Grill-Spector and Malach, 2004), and macaque (Richmond et al., 1990; Rolls, 1991; Logothetis and Sheinberg, 1996; Keysers et al., 2001; Hung et al., 2005; Kiani et al., 2005; Connor et al., 2007) and human (Allison et al., 1999; Privman et al., 2007; Liu et al., 2009) neurophysiological recordings, presentation of each image part (Figure 2B) or the whole image (Figure 2C) elicited a strong selective response commencing approximately 100 ms after stimulus onset (see also Figure 3). For long SOAs (e.g. 250 ms, Figure 2D5, E5), the sequential parts elicited distinct responses that were largely independent. As the SOA became shorter, these responses to the two parts overlapped and revealed evidence of interactions (Figure 2D1-4,E1-4). Responses to asynchronous presentations were different from those to the whole two-part image (permutation test, $p=0.0001$), suggesting that both spatial and temporal features in the stimulus influence the electrode's responses. Moreover, there was a striking dependency on the temporal order with which the two parts were presented (compare Figure 2D versus 2E). At long SOAs, dependence on temporal order is trivial given the independence of the responses to each image part. Yet, such temporal order dependencies were evident even at the shortest SOA that we tested (17 ms, permutation test, p=0.002, Figure 2D1 versus E1).

The influence of stimulus presentation dynamics on the physiological responses could reflect spatiotemporal features sensitive to input timing and to presentation order. To evaluate the relative influence of SOA and presentation order, we constructed in Figure 4A a matrix of pairwise root-mean-square-error (RMSE, Eq. 11) differences between all the responses shown in Figure 2C-E. This comparison matrix showed smaller differences between conditions in which the two parts were presented in the same order than when they were presented in opposite orders (see Figures 4B-C for other examples). To quantify this observation, we calculated an order-tuning index (OTI, Eq. 14). The OTI ranged from -1 (differences *between* orders negligible compared to SOA-dependent differences *within* orders) to 0 (differences within orders as large as differences between orders) to 1 (differences within orders negligible compared to differences between orders). The OTI for the example electrode in Figure 1 was 0.41. There were 692 responses from 173 electrodes located in ventral visual regions (692 = 173 electrodes x 4 two-part images, Table 1). We calculated OTI values for the 221 of these 692 responses that were visually responsive to the whole image (e.g. Figure 2C) and to both constituent halves (e.g. Figure 2B, Methods). The average comparison matrix revealed a clear asymmetry depending on whether differences were computed across versus within orders (Figure 4D). All OTI values were positive (mean±SD = 0.32±0.09, Figure 4E). A variant of the OTI, which considered cross-order differences only up to 33ms SOA, also showed a positive bias (OTI'=0.27±0.12, mean±SD). To minimize the possibility that non-visual signals could influence the results, we repeated the OTI computations considering only data between 50 and 200 ms after stimulus onset (Figure 4F-G). Within this window, the OTI values were even more positive (OTI=0.40±0.12; OTI' = 0.39±0.14). The positive order tuning indices demonstrate that these visual responses were more sensitive to temporal disruptions that reversed the order of part presentations than to disruptions that preserved order.

We sought to quantitatively describe the dynamic interactions between asynchronous halves that gave rise to this order sensitivity by considering a series of linear models and evaluating the probability that the physiological data could arise from each model (Methods). To avoid cases where such models could trivially explain the data, we restricted the analyses to those SOAs that revealed the strongest interactions between parts (17 ms, 33 ms and 50 ms) and we focused on 111 responses at 54 electrodes showing significant temporal order sensitivity at 17 ms SOA (permutation test, p<0.05, e.g. Figure 2D1, E1). The responses that did not show such temporal order sensitivity are described in Figure 5. The models showed an even better performance in explaining the order-insensitive responses; we focus the rest of the manuscript on describing the responses that are more challenging to explain, namely those that showed order sensitivity.

Waveforms elicited by whole images (Eq. 1) poorly modeled the data, describing only 8% of responses, which demonstrated that short temporal asynchrony significantly changed the physiological responses. A model where one image half might dominate and be entirely responsible for driving the observed response (Gawne and Martin, 2002; Zoccolan et al., 2007; Agam et al., 2010) (Eq. 2) explained 34% of the data. We next considered a weighted linear combination of both halves' responses (Baker et al., 2002; Zoccolan et al., 2007; Agam et al., 2010; Baeck et al., 2013), shifted to reflect SOA (Eq. 3, "simple linear model"). Figure 6A-D shows an example of an electrode located in the left fusiform gyrus where the data were well fit by this model (p=0.21; compare orange traces versus green traces), but not by either of the individual image halves.

The simple linear model provided a good description of the data for 44% of the responses. Yet, there were electrodes where the interactions between image halves could not be explained by this model (e.g. Figure 6E-H). Inspired by the temporal sensitivity documented in Figures 2 and 4, we extended Eq. 3 by introducing a general linear model endowed with the flexibility to reflect relative timing by taking into account the SOA *and* presentation order (Eq. 4). Incorporating relative timing allowed the general linear model to describe 62% of the responses, an increase of 41% over the simple linear model (e.g. Figure 7). The increased explanatory power was not merely due to the addition of free parameters: the general linear model was found to be a significantly better description of the data for 72% of the responses, according to the Akaike Information Criterion ($\Delta AICc > 4$, Methods). Further evidence for the importance of relative timing was provided by a final pair of models that evaluated the similarity of waveforms as SOA increased beyond 17ms. The simpler such model (Eq. 5) used coefficients that depended only on temporal order, and described 48% of responses (e.g. Figure 6E-H). Allowing coefficients to also depend on SOA (Eq. 6) did not describe any additional responses.

The presentation of different parts in asynchronous fashion generates spatiotemporal features that are absent when the parts are presented simultaneously. These features may lead to electrophysiological responses that differ between the two presentation orders, irrespective of the parts' shapes. For example, inhomogeneities between the top and bottom parts of an electrode's receptive field or apparent motion signals could lead to distinct neural signals depending on which part is presented first. To evaluate whether such spatiotemporal features, irrespective of the parts' shapes, could explain the signals described above, we compared responses to two distinct sets of parts presented under the same order and SOA (Eq. 7). Figure 8A shows the responses of the same electrode from Figure 2, comparing the responses to part b3 followed 17 ms later by t3 (green, same as Fig 1D1) versus part b4 followed 17 ms later by t4 (purple). Figure 8B shows the corresponding responses when the presentation order was reversed.
In all, only 8% of the order-sensitive responses could be described by the model described by Eq. 7. While spatiotemporal features that depend on presentation order are likely to contribute to the responses documented here, the differences between the traces in each panel in Figure 8 suggest that the responses cannot be purely explained in terms of features that are independent of the parts' shapes.

We parceled each subject's brain into 75 anatomical regions based on pre-operative MR and post-operative CT images (Methods). We focused on the five regions along the ventral visual stream with more than five order-sensitive visual responses: occipital pole, inferior occipital gyrus, inferior temporal gyrus, middle occipital gyrus and fusiform gyrus. We compared the performance of the simple linear model (Eq. 3) versus the short SOA model (Eq. 5), evaluated at 33ms and 50ms SOA. The short SOA model performed significantly worse at the occipital pole (Figure 9, chi-square test, p=0.02), but no such differences were observed in the four higher visual areas (chi-square tests, p>0.2). Performance of the short SOA models (Eq. 5-6) varied significantly by region (chi-square test, p=0.02), explaining 31% of responses at the occipital pole and 83% of responses in inferior temporal gyrus. By contrast, performance of the part-based linear models (Eq. 3-4) remained relatively consistent across regions (chi-square tests, p>0.65).

During the electrophysiology experiments, subjects focused on recognizing the two-part images regardless of the presentation asynchrony. To evaluate whether it is possible to detect asynchronous presentation, we conducted a separate behavioral experiment (without physiological recordings) under identical stimulus presentation conditions but asking subjects to determine whether or not the parts were presented synchronously. The behavioral data suggested that subjects were able to identify asynchrony with SOAs of 50 or 100 ms but not with SOAs of 17 ms (Figure 10). Extrapolating from these results, it seems plausible that the SOA=17 ms condition was perceptually quite close to the SOA=0 ms condition during the physiological recordings.

**Discussion**

The current study investigated the neural representation of image parts that were asynchronously presented with intervals ranging from 17 to 250 ms in a task where subjects had to put together the two components to form a whole. Examining intracranial field potentials recorded along the human ventral visual stream, long SOAs of 100 to 250 ms led to largely independent responses to each separate image part (e.g. Figure 2, D5 and E5). By contrast, the responses revealed interactions between the parts at shorter asynchrony values. The neural signals reflecting the integration of the two parts were sensitive to the asynchrony of stimulus presentation, even at an SOA as short as 17 ms (e.g. compare Figure 2C versus 2D1). Moreover, which part was presented first strongly influenced the ensuing response (e.g. compare Figure 2, D1 versus E1).

Spatial context is known to modulate responses throughout visual cortex from early visual areas (Zipser et al., 1996; Bair et al., 2003; Angelucci and Bressloff, 2006; Allman et al., 1985) all the way to intermediate and higher visual areas (Missal et al., 1999; Zoccolan et al., 2007; Reynolds et al., 1999; Chelazzi et al., 1993; Agam et al., 2010; Baeck et al., 2013). In early visual cortex, adding visual stimulation in the surround of a neuron's receptive field typically inhibits the responses to the center stimulus (Sceniak et al., 1999; Allman et al., 1985). Additionally, recordings in visual areas V4 and IT have shown that presenting two or more stimuli (sometimes referred to as "clutter" in the context of visual search tasks) within a neuron's receptive field typically leads to a reduction in the firing rate to the preferred stimulus (Missal et al., 1999; Zoccolan et al., 2007; Reynolds et al., 1999; Chelazzi et al., 1993). In the absence of attention, the responses to object pairs have been successfully described by linear weighted models similar to Equation 3 from the current manuscript (with SOA=0) (Reynolds et al., 1999; Baeck et al., 2013; Zoccolan et al., 2007). The success of these linear weighted models in describing many (but not all, Figure 5) of the responses in the current study extends the notion of linear combinations to the time domain.

A recent study from our lab showed that the weighted sums were biased towards the response to the preferred objects in situations where two random objects were disconnected, independent, simultaneously presented and there was no behavioral integration required during the task (Agam et al., 2010), similar to the studies of (Zoccolan et al., 2007; Gawne and Martin, 2002) and one of the conditions in (Baeck et al., 2013). The current observation that a single part model (Equation 2) was a good description only for a few of the responses (Figure 5) suggests that the distance between parts and/or the task's requirement to incorporate information from the two parts may play an important role in shaping the responses to spatial context along the ventral visual stream.

In addition to spatial context, temporal context can also influence physiological responses in early visual areas (Nelson, 1991; Vinje and Gallant, 2002; Bair et al., 2003; Benucci et al., 2009). Less is known about how temporal context modulates neurophyisological signals in higher visual areas. Neuroimaging signals have shown dependence on stimulus temporal order over scales of hundreds of ms to seconds in higher visual areas (Hasson et al., 2008) and several studies have documented how recognition at the behavioral level can be influenced by asynchronous presentation of visual stimuli (Clifford et al., 2004; Eagleman et al., 2004; Singer and Sheinberg, 2006; Anaki et al., 2007). The current study demonstrates that temporal context can significantly affect physiological responses throughout the human ventral visual stream, even within as little as 17 ms. In several cases, the effects of temporal context could be accounted for by adding a time shift to the weighted linear models (Equations 3-4, Figure 5, Figure 6A-D). However, many other responses manifested a strong sensitivity to the order in which the stimuli were presented (Figure 2, Figure 6E-H, Figure 5) that could not be accounted by the time-shifted weighted linear models.

We considered whether the current results could be explained in terms of differential eye movements, adaptation effects or masking effects. Given that stimuli and asynchrony values were presented in random order and with short asynchrony, it seems difficult to explain the results based on differential eye movements between the whole condition and the 17 ms SOA conditions or between the two 17 ms asynchrony conditions with different orders. Even if subjects were to make distinct saccades triggered and dictated by the stimulus order, it is unclear whether such saccades could be rapid enough to explain differences in the physiological responses during the initial 200 ms. The sensitivity to small asynchrony intervals also argues against an interpretation of the data presented here based on adaptation over long time scales. Short intervals are typical in visual masking studies. A mask presented within tens of ms of a visual stimulus (either before or after) can strongly influence neurophysiological responses as well as visual perception (Kovacs et al., 1995; Macknik, 2006; Felsten and Wasserman, 1980; Rolls et al., 1999). It seems unlikely that the effects documented here can be ascribed to masking given that (i) there is no clear masking effect at the behavioral level (Figure 10), (ii) in most visual masking studies, the mask spatially overlaps the primary stimulus, and (iii) the compatible content and adjacent peri-foveal organization of the images argue against paracontrast or metacontrast masking (Alpern, 1953).

The origin of the sensitivity to relative timing reported here is not clear. Distinct spatiotemporal features that depend on stimulus order elicited distinct responses even as early as the occipital pole. Consistent with potential early origins, prior work has found that neurons in primary visual cortex show modulation in their responses when stimuli inside or outside their receptive fields are presented in temporal proximity (Nelson, 1991; Bair et al., 2003; Benucci et al., 2009). The dependence on relative stimulus timing described in the current study is particularly intriguing in light of theoretical and experimental studies proposing that a robust and efficient representation of information can be encoded in the temporal order with which neurons fire action potentials (Hopfield, 1995; vanRullen and Thorpe, 2002; Gollisch and Meister, 2008) The current observations suggest that spatiotemporal interactions persist even at the highest levels of visual processing within the ventral pathway, and with dynamics on the scale of tens of milliseconds..

# Figures

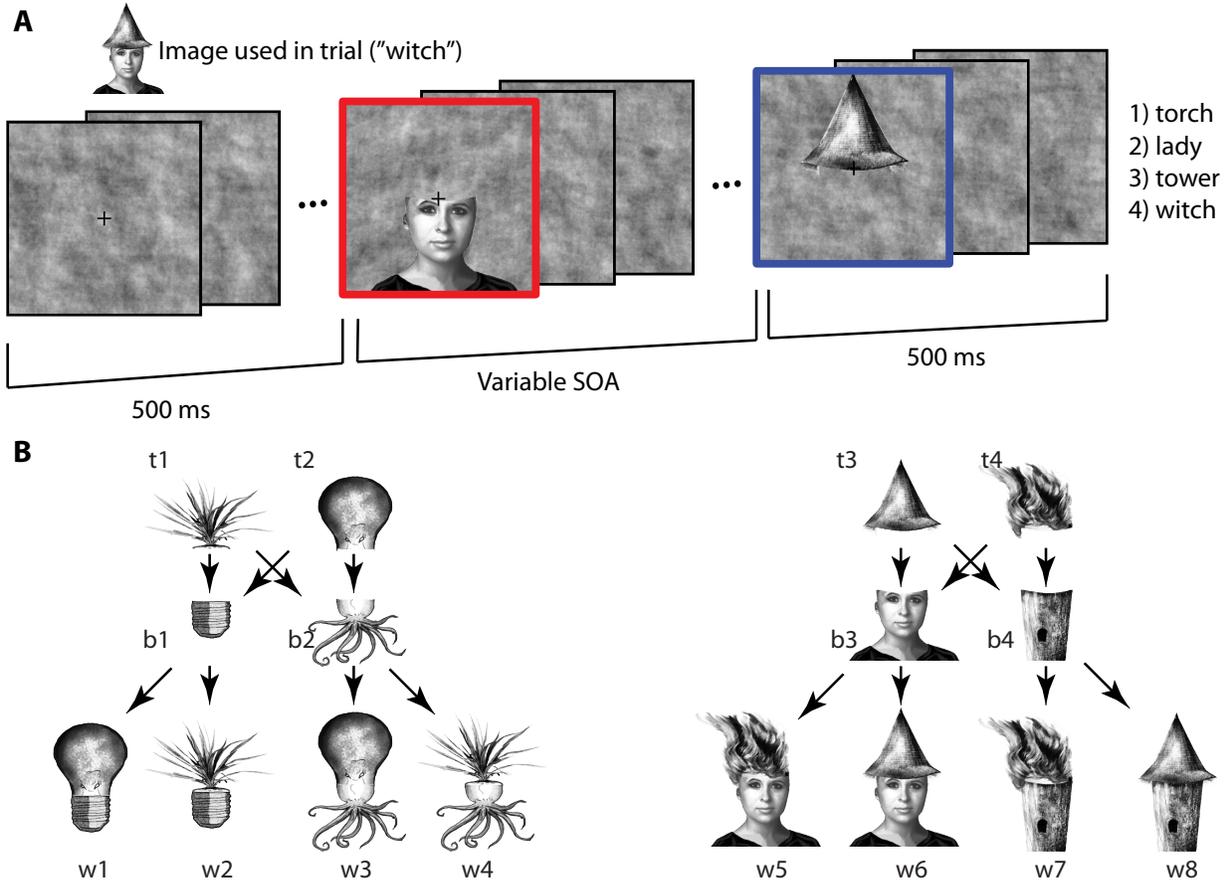

**Figure 1.** Trial structure and images used.
(**A**) Subjects identified which of 4 two-part images was shown in a 4-alternative forced-choice task (Methods). Each part was flashed for 17 ms in the midst of low-contrast noise. Parts were separated by a stimulus onset asynchrony (SOA) ranging from 0 to 250 ms. The order of presentation of SOAs, images and parts was randomized across trials. Frames did not have colored borders in the actual experiment (shown here in relation to Figure 2).
(**B**) The two sets of images used. Each set contained two bottom part and two top parts that were combined to form four possible whole images..Each part was used in two images. w1 ("light bulb") = b1+t2; w2 ("houseplant") = b1+t1; w3 ("octopus") = b2+t2; w4 ("turnip") = b2+t1. w5 ("lady") = b3+t4; w6 ("witch") = b3+t3; w7 ("torch") = b4+t4; w8 ("tower") = b4+t3.

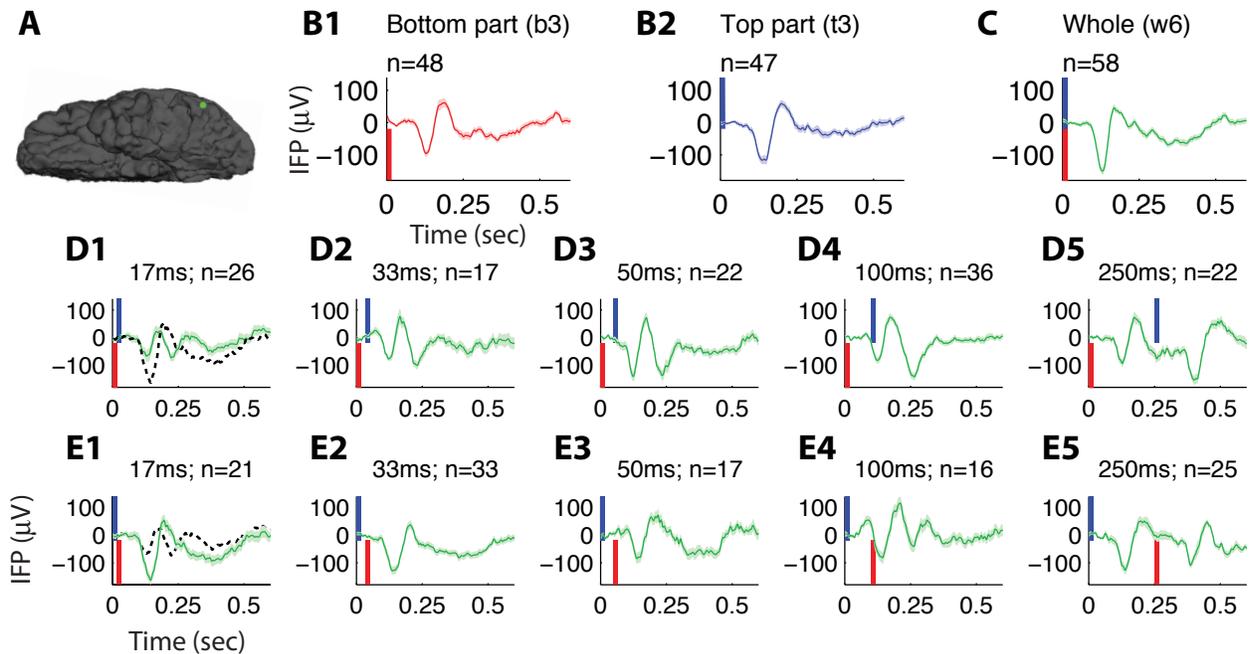

**Figure 2**. Example physiological responses.
(**A**) Example electrode located in the left fusiform gyrus. (**B**) Intracranial field potential responses to the two parts from Figure 1A (b3, t3) when presented independently (red = b3, blue = t3), aligned to image part onset and averaged over 48 and 47 repetitions respectively. Shaded areas denote SEM. (**C**) Responses to the whole image (w6). (**D-E**) Responses at increasing SOA values, with the bottom part shown first (**D**) or last (**E**). Responses are aligned to the first part onset. The red (blue) rectangle denotes the onset and duration of the bottom (top) part. The dotted line in **D1** shows the response in the opposite-order condition (**E1**) for comparison purposes, and the dashed line in **E1** shows the response from **D1**. Even at the shortest non-zero SOA (17 ms), responses differed depending on part order (cf. **D1** versus **E1**, *p*=0.002, permutation test).

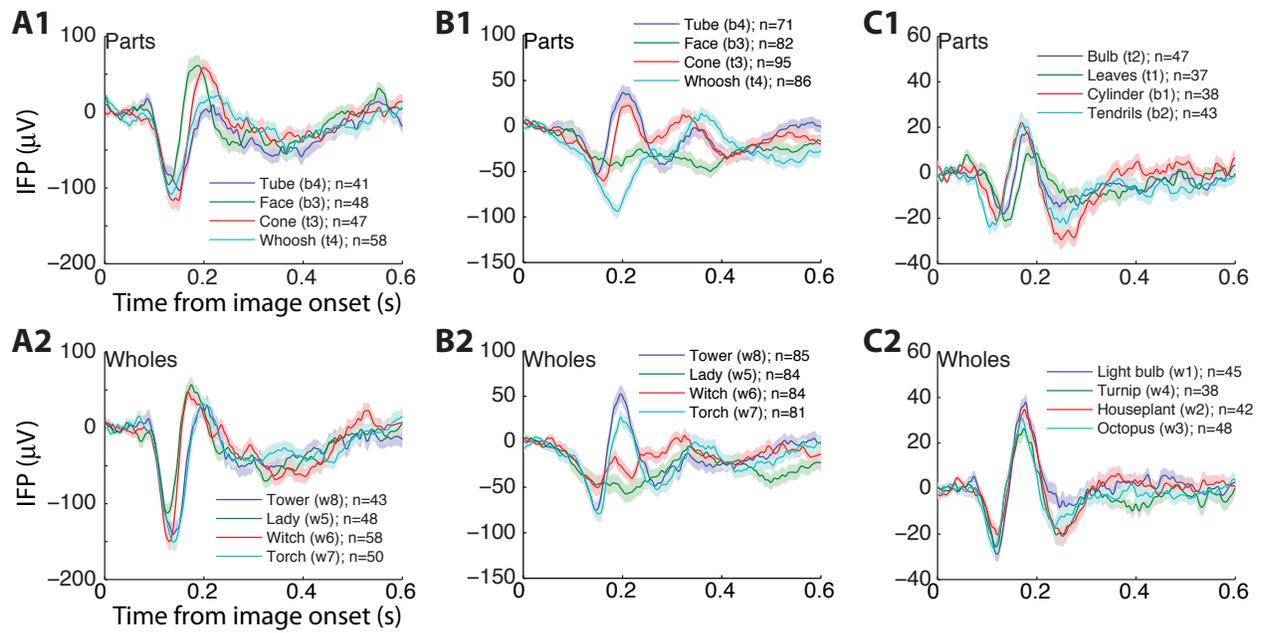

**Figure 3.** Responses to all images and image parts for three example electrodes. Responses to all parts (top row) and wholes (bottom row) from three electrodes located in the left fusiform gyrus (**A** & **B**) and left middle occipital gyrus (**C**). The format and conventions are the same as those in Figure 2B-C. The electrode in part A is the same as the one shown in Figure 2 (the responses to parts b3 and t3 in A1 were shown in Figure 2B and the responses to w6 in A2 were shown in Figure 2C).

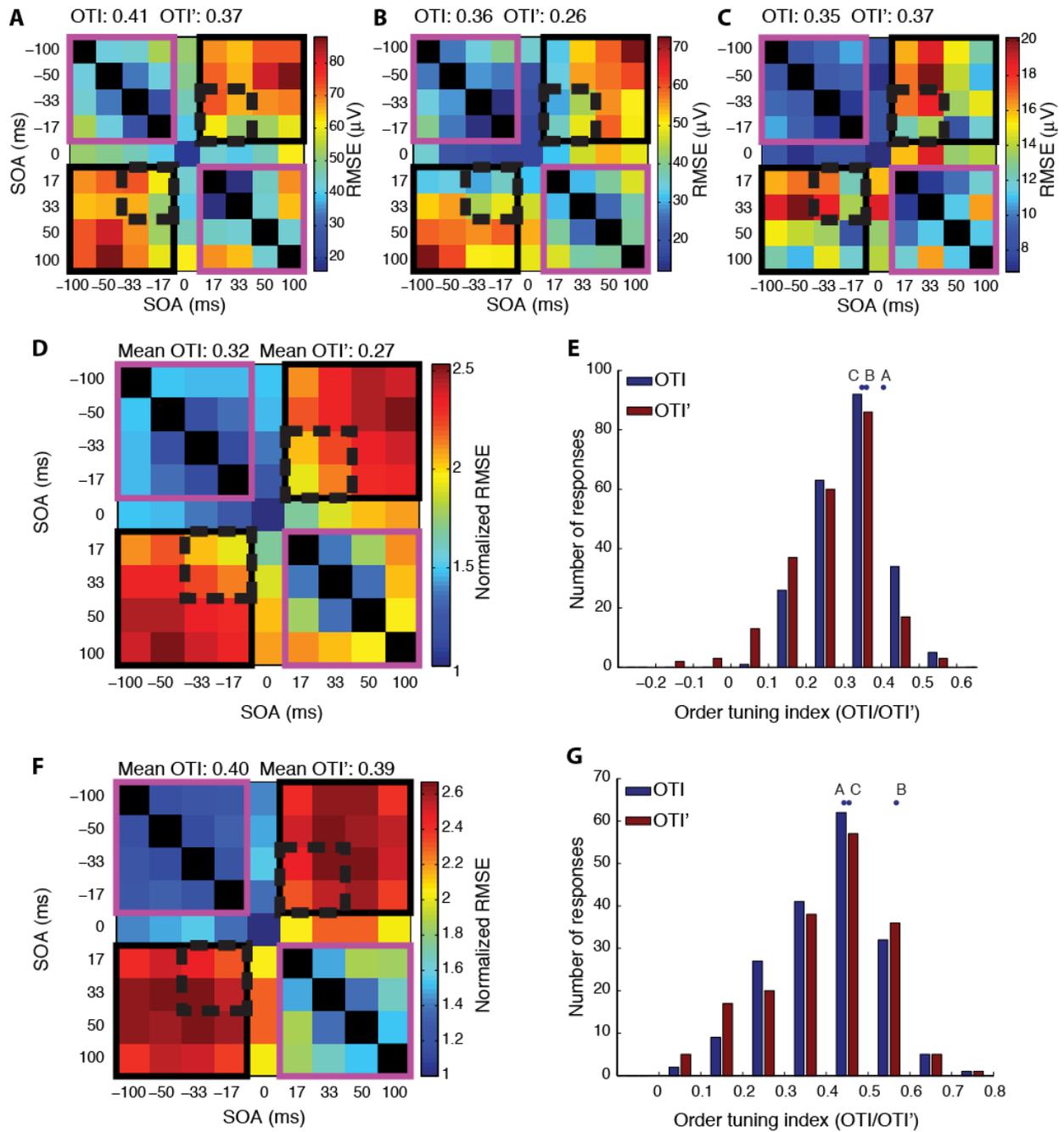

**Figure 4.** Comparison of responses across orders and SOAs.
(**A-C**) For each of the three example electrodes in Figure 3, we computed a comparison matrix contrasting responses at different SOAs and presentation orders. Entry *i, j* in this matrix represents the root-mean-square error (Eq. 11) between mean responses elicited by trials with SOAs given by *i* and *j* (see color scale on the right). The root mean square was not computed along the diagonal, which is shown as black squares. The order tuning index (OTI) was calculated from the mean difference between responses with different temporal orders (solid black windows) minus the mean difference between responses with the same temporal order (pink windows, Eq. 14). In the modified order tuning index (OTI'), we replaced the solid black window with the dashed black window (Eq. 16). Positive OTIs indicate that differences between SOA conditions that preserve order are smaller than those between SOA conditions that do not.
(**D**) Summary comparison matrix contrasting responses at different SOAs and presentation orders (n=221 responses from 78 electrodes, Methods). Here the RMSE is normalized for each electrode before averaging.
(**E**) Distribution of OTI (blue) and OTI' (red). Bin size = 0.1. OTI values for the three example electrodes are indicated with circles.
(**F-G**) Same analysis as in part D-E except using the interval between 50 and 200 ms after stimulus onset (n=179 visual responses). Mean OTI=0.40±0.12; mean OTI' = 0.39±0.14.

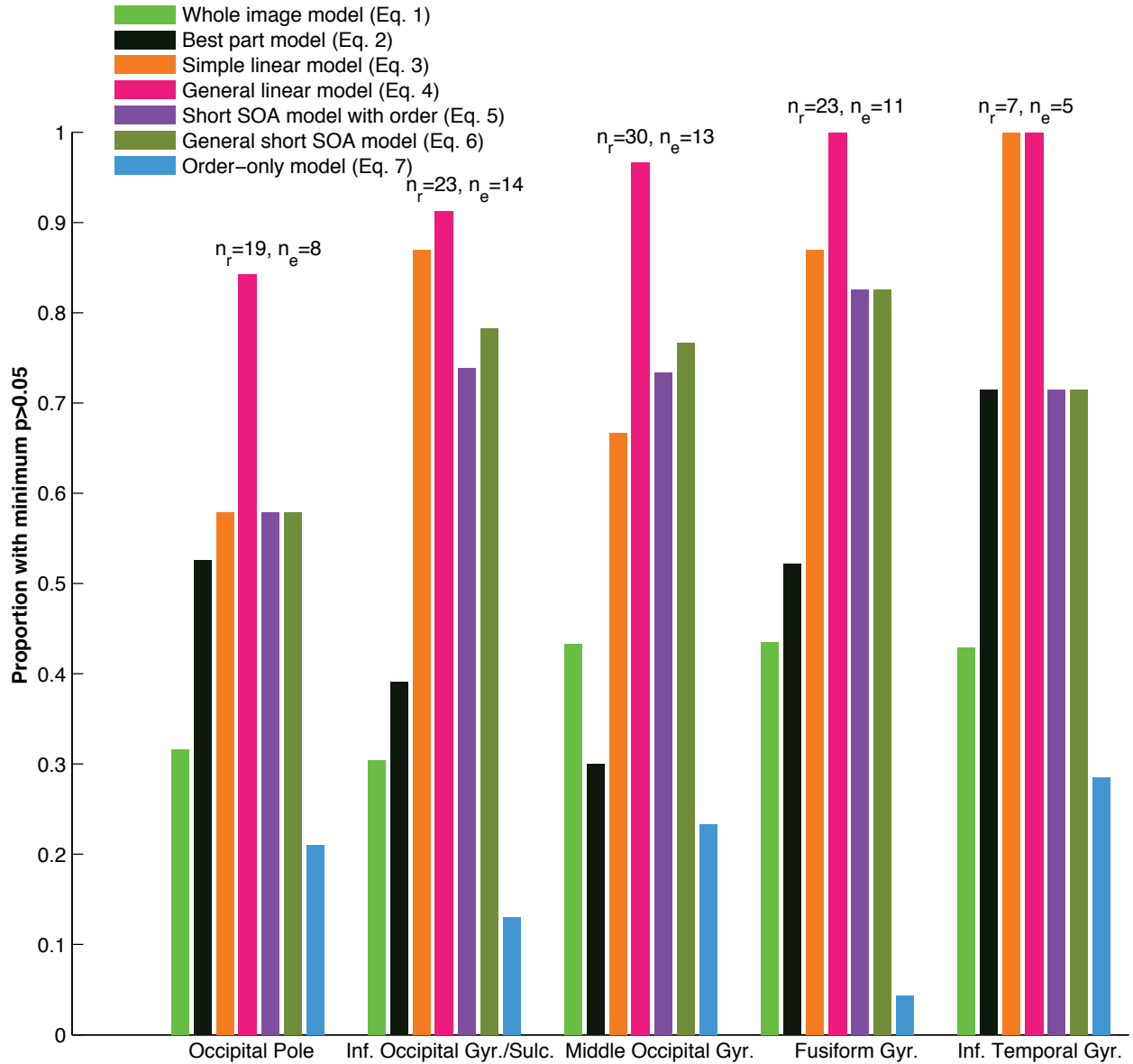

**Figure 5.** Comparison among quantitative models for order-insensitive responses. Performance of 7 different models at the five visual surface regions with >5 responses, considering only order-insensitive responses (i.e., responses that did not show a significant difference between the two presentation orders at 17ms). For each region, $n_r$ indicates the number of responses and $n_e$ the number of electrodes considered. For each model, in each region, the proportion of responses characterized with probability p>0.05 is shown. Here the models are compared at SOA=±33 and ±50 ms (and not at SOA=±17ms because Eq. 5-6 are not defined at this SOA). The models are:

"Whole image model" (Eq. 1, light green). The response at each SOA is compared to the response to the whole image.

"Best part model" (Eq. 2, black). Whichever image part response gives a higher probability across all SOAs is compared to the response at each SOA.

"Simple linear model" (Eq. 3, orange). One of the models considered in Figure 8. One coefficient is fit to the response to each image part, and the response at each SOA is compared to the weighted sum (2 total coefficients).

"General linear model" (Eq. 4, magenta). One coefficient is fit to the response to each image part at each SOA, and the response at each SOA is compared to the weighted sum (8 total coefficients).

"Short SOA model with order" (Eq. 5, purple). One of the models considered in Figure 8. The response at each SOA is compared to the response at 17ms SOA, scaled by a coefficient fit to trials with the same order (2 total coefficients).

"General short SOA model" (Eq. 6, dark green). The response at each SOA is compared to the response at 17ms SOA, scaled by a coefficient fit to that SOA and order (4 total coefficients).

"Order-only model" (Eq. 7, blue). The response at each SOA is compared to the response produced by the two parts *not* used in the image in question, with the same SOA and order.

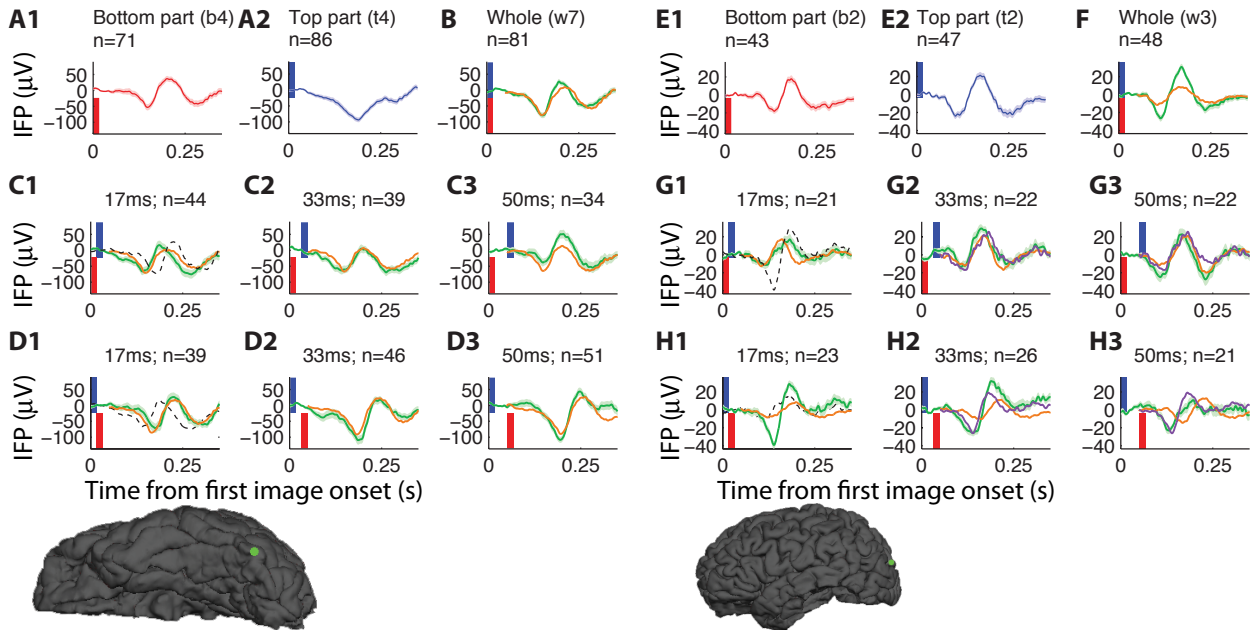

**Figure 6.** Examples of quantitative modeling of the physiological responses. Panels A-D show the responses recorded from an electrode in the left fusiform gyrus upon presentation of the w7 stimulus. Individual panels use the same conventions as Figure 2B-E. The responses of this electrode to all parts and wholes is shown in Figure 3B. Panels E-H show the responses recorded from an electrode in the left middle occipital gyrus upon presentation of the w3 stimulus. The responses of this electrode to all parts and wholes is shown in Figure 3C. The orange lines in C-D and G-H show fits from a linear model that aims to predict the responses to the images at different SOA levels from the constituent parts (simple linear model, Eq. 3, Methods). This model provided a good fit to the data in C-D (p=0.21, Methods) but not in G-H (p=0.007). Purple lines in G-H show the fits of a model that compares responses at 33ms and 50ms SOA to responses at 17 ms SOA, with a coefficient that varies depending on order (short-SOA model with order, Eq. 5, Methods). This model described the data well (p=0.20).

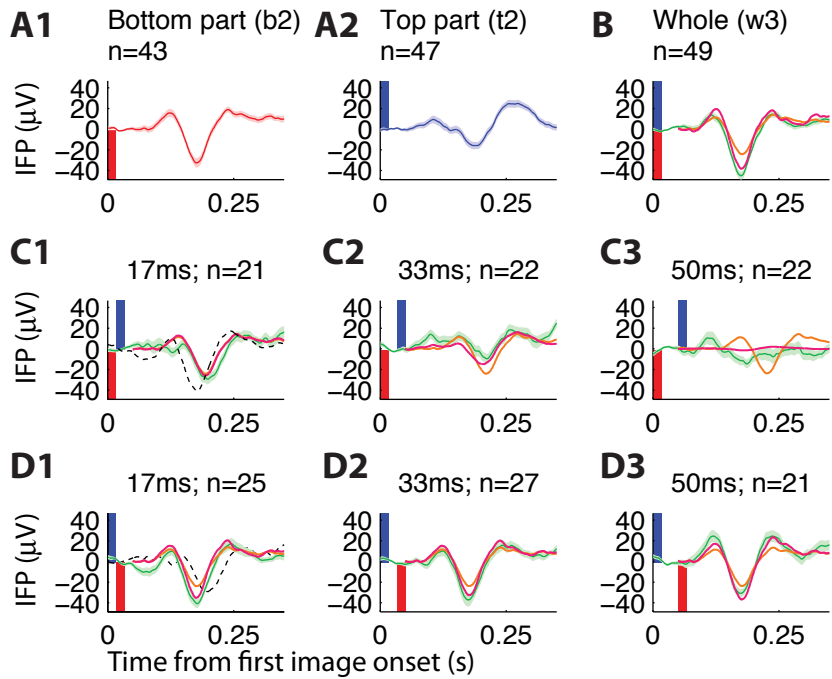

**Figure 7.** Example electrode with responses described by the general linear model. Responses from an electrode in left fusiform gyrus (see inset for electrode location) upon presentation of the w3 stimulus. Conventions are as in Figure 6. While the simple linear model (Eq. 3, orange) failed to describe the data (p=0.02), the general linear model (Eq. 4, magenta) was able to capture the different waveforms as SOA and order varied (p=0.28).

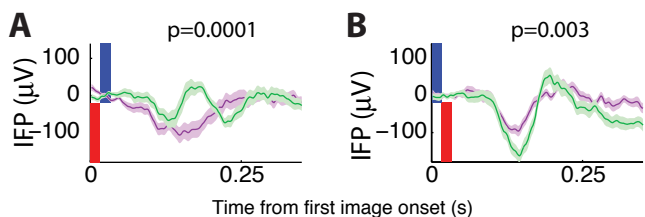

**Figure 8.** Example electrode comparing different parts presented with the same order. For the same electrode shown in Figure 2, we compare the responses elicited by presentation with 17 ms SOA of parts b3+t3 (green) versus parts b4+t4 (purple) when the bottom part was presented first (**A**) or last (**B**). The presentation order was the same for the green versus purple curves in each subplot; the differences between the curves demonstrate that part identity was critical to the physiological responses.

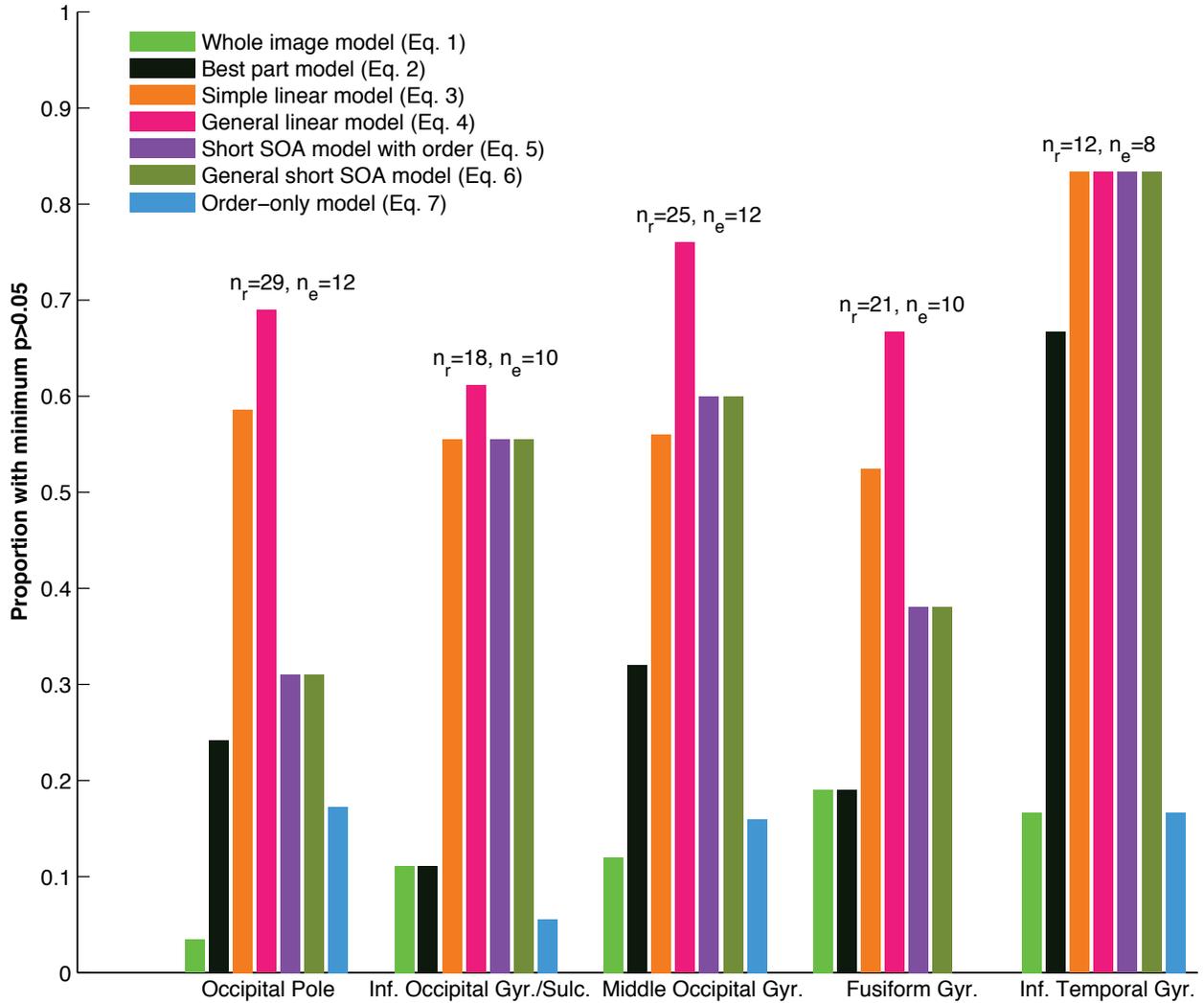

**Figure 9.** Comparison of all models for order sensitive responses. Performance of 7 different models at the five visual regions with >5 responses, considering only order-sensitive responses (see Methods for description of each model). Conventions are otherwise the same as in Figure 5.

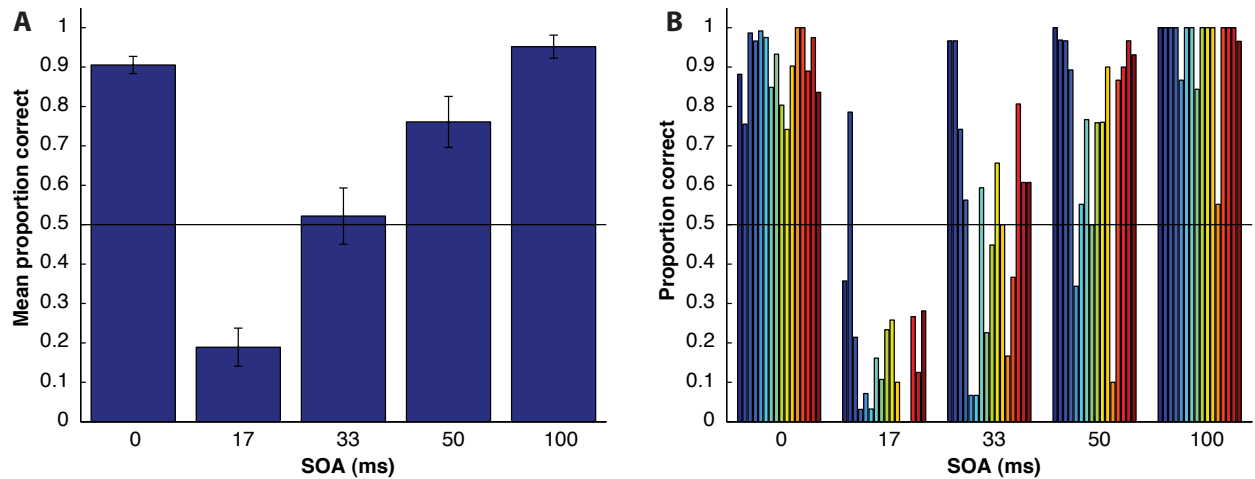

**Figure 10.** Psychophysics experiment determining whether subjects can determine asynchrony. Stimuli and presentation parameters were identical to those used in the main experiment, but subjects were asked to indicate whether or not the two parts were presented simultaneously (2-alternative forced choice). Horizontal black lines indicate chance-level performance. (**A**) Mean performance (±S.E.M.) across all 16 subjects. (**B**) Each color denotes a different subject. There were no physiological recordings during this variant of the experiment.

**Table 1: Electrode/response properties and counts by brain region**

| Region | N Electrodes | N responses to whole | N responses to both halves | N order sensitive responses | N Eq. 3 | N Eq. 5 |
|---|---|---|---|---|---|---|
| Anterior transverse collateral sulcus | 3 | 7 (58%) | 4 (57%) | 0 (0%) | 0 (0%) | 0 (0%) |
| Fusiform gyrus | 24 | 54 (56%) | 44 (81%) | 21 (48%) | 11 (25%) | 8 (18%) |
| Inferior occipital gyrus/sulcus | 25 | 55 (55%) | 41 (75%) | 18 (44%) | 10 (24%) | 10 (24%) |
| Occipital pole | 35 | 70 (50%) | 48 (69%) | 29 (60%) | 17 (35%) | 9 (19%) |
| Lateral occipitotemporal sulcus | 2 | 4 (50%) | 4 (100%) | 2 (50%) | 0 (0%) | 1 (25%) |
| Middle occipital/lunate sulcus | 5 | 8 (40%) | 6 (75%) | 4 (67%) | 0 (0%) | 0 (0%) |
| Middle occipital gyrus | 40 | 63 (39%) | 55 (87%) | 25 (45%) | 14 (25%) | 15 (27%) |
| Inferior temporal gyrus | 39 | 31 (20%) | 19 (61%) | 12 (63%) | 10 (53%) | 10 (53%) |
| **Visual Cortex Totals** | **173** | **292** | **221** | **111** | **62** | **53** |
| Intraparietal sulcus | 6 | 3 | 0 | 0 | | |
| Parahippocampal gyrus | 22 | 6 | 1 | 0 | | |
| Superior temporal sulcus | 11 | 2 | 2 | 0 | | |
| Cuneus | 23 | 4 | 2 | 0 | | |
| Superior occipital gyrus | 7 | 1 | 0 | 0 | | |
| Precentral gyrus | 9 | 1 | 0 | 0 | | |
| Middle temporal gyrus | 82 | 7 | 3 | 0 | | |
| Supramarginal gyrus | 47 | 4 | 1 | 0 | | |
| Lingual gyrus | 39 | 3 | 1 | 0 | | |
| Depth electrodes | 26 | 2 | 0 | 0 | | |
| Subcentral gyrus/sulcus | 14 | 1 | 0 | 0 | | |
| Angular gyrus | 28 | 2 | 1 | 0 | | |
| Lateral superior temporal gyrus | 49 | 3 | 0 | 0 | | |
| Planum temporale | 5 | 0 | 0 | 0 | | |
| Postcentral gyrus | 14 | 0 | 0 | 0 | | |
| Postcentral sulcus | 3 | 0 | 0 | 0 | | |
| Temporal pole | 18 | 0 | 0 | 0 | | |
| Short insular gyri | 1 | 0 | 0 | 0 | | |
| Operculum | 6 | 0 | 0 | 0 | | |
| Posterior ventral cingulate | 1 | 0 | 0 | 0 | | |

| Region | | | | |
|---|---|---|---|---|
| Collateral/lingual sulcus | 4 | 0 | 0 | 0 |
| Inferior temporal sulcus | 5 | 0 | 0 | 0 |
| Precuneus | 8 | 0 | 0 | 0 |
| Superior/transverse occipital sulcus | 2 | 0 | 0 | 0 |
| Unclassified | 1 | 0 | 0 | 0 |
| Superior parietal gyrus | 4 | 0 | 0 | 0 |
| Calcarine sulcus | 1 | 0 | 0 | 0 |
| Triangular inferior frontal gyrus | 4 | 0 | 0 | 0 |
| Parieto-occipital sulcus | 1 | 0 | 0 | 0 |
| Middle frontal gyrus | 2 | 0 | 0 | 0 |
| Hippocampus | 2 | 0 | 0 | 0 |
| **Totals** | **618** | **331** | **232** | **111** |

**Table 1:** Summary of electrode locations and response properties. We parceled the brain into 75 regions (Methods) and report the number of electrodes in each region, the number of responses to whole images (percentage with respect to total responses in parentheses), the number of responses to both image halves (percentage with respect to responses to whole images in parentheses), the number of order-sensitive responses at 17 ms (percentage with respect to responses to both halves in parentheses), the number of responses described by the simple linear model (Eq. 3, percentage with respect to responses to both halves in parentheses) and the short SOA model with order (Eq. 5, percentage with respect to responses to both image halves in parentheses).